\documentclass[10pt, conference, compsocconf]{IEEEtran}
\IEEEoverridecommandlockouts

\ifCLASSINFOpdf
\else
\fi
\usepackage{graphicx}
\usepackage{amsmath}
\usepackage{subfigure}
\usepackage{cite}
\usepackage{comment}

\begin{document}

\title{Advanced coding schemes against jamming in telecommand links}



\author{\IEEEauthorblockN{M. Baldi\IEEEauthorrefmark{1}, M. Bianchi\IEEEauthorrefmark{1}, F. Chiaraluce\IEEEauthorrefmark{1}, R. Garello\IEEEauthorrefmark{2}, N. Maturo\IEEEauthorrefmark{1}, I. Aguilar Sanchez\IEEEauthorrefmark{3}, S. Cioni\IEEEauthorrefmark{3}}
\IEEEauthorblockA{\IEEEauthorrefmark{1}DII, Universit\`a Politecnica delle Marche \\
Ancona, Italy \\
Email: \{m.baldi, m.bianchi, f.chiaraluce, n.maturo\}@univpm.it }
\IEEEauthorblockA{\IEEEauthorrefmark{2}DET, Politecnico di Torino \\
Torino, Italy \\
Email: garello@polito.it}
\IEEEauthorblockA{\IEEEauthorrefmark{3}TEC-ETC, ESA-ESTEC\\
Noordwijk, The Netherlands\\
Email: \{ignacio.aguilar.sanchez, stefano.cioni\}@esa.int}
\thanks{Copyright (c) 2013 IEEE. Personal use of this material is permitted. However, permission to use this material for any other purposes must be obtained from the IEEE by sending a request to pubs-permissions@ieee.org.

This work was supported in part by the ESA Contract No: 4000106268: Advanced Coding Schemes for Direct Sequence Spread Spectrum Telecommand Links and in part by the MIUR project ``ESCAPADE'' (Grant RBFR105NLC) under the ``FIRB - Futuro in Ricerca 2010'' funding program.}
}

\maketitle

\begin{abstract}
The aim of this paper is to study the performance of some coding schemes recently proposed for updating the TC channel coding standard for space applications, in the presence of jamming. Besides low-density parity-check codes, that appear as the most eligible candidates, we also consider other solutions based on parallel turbo codes and extended BCH codes. We show that all these schemes offer very good performance, which approaches the theoretical limits achievable.
\end{abstract}


\IEEEpeerreviewmaketitle

\section{Introduction}
\label{sec:one}

\IEEEPARstart{S}pace missions can be impaired by intentional or unintentional jamming. Such a threat is particularly dangerous for telecommands (TC), since the success of a mission may be compromised because of the denial of signal reception by the satellite. It is well known that to counter the jamming threat, error correcting codes can be used jointly with direct sequence spread spectrum. 
This topic has been investigated in previous literature, but rarely taking into account the peculiarities of the TC space link. As a consequence, in current standards or recommendations on TC space links, only weak countermeasures are included, that do not appear adequate to face the increasing skill of malicious attacks. Whilst for the spreading technique a relevant advance is brought by the introduction of long cryptographic pseudo-noise sequences \cite{Thales2011}, the discussion is quite open as regards possible usage of new coding techniques.

As a matter of fact, the only error correcting code currently included in the standards and recommendations for TC applications \cite{ECSS2008}, \cite{CCSDS2010} is the BCH code with dimension $k = 56$ and length $n = 63$ exploiting hard-decision decoding. The performance of this code in the presence of jamming is generally not good and it is possible to verify that significant losses appear even when an interleaver (not currently included in the standard) is employed. Actually, the performance of the BCH$(63, 56)$ code is unsatisfactory even when considered on the additive white Gaussian noise (AWGN) channel. For this reason a number of new proposals have been formulated with emphasis on binary and non-binary low-density parity-check (LDPC) codes \cite{CCSDS2012}, \cite{NASA/JPL2012}, \cite{Liva2012}.
Since good correction capability and short frames are needed, the most recent proposals consider codes with rate $R_c = 1/2$ and $k = 64$, $128$ or $256$.
Using non-binary LDPC codes allows to improve on their binary counterparts \cite{Costantini2012}: as an example, a non-binary LDPC$(128, 64)$ code at a codeword error rate (CER) of about $10^{-4}$ gains roughly $1$ dB over the binary LDPC$(128, 64)$ code.
The performance of these codes over jamming channels has not yet been investigated.

In this paper we consider different types of jamming signals, namely: pulsed jamming, continuous wave (CW) jamming and pseudo-noise (PN) jamming. We also study the impact of jammer state information (JSI), clipping and interleaving,  under typical TC application constraints.
In order to assess how far the performance of the considered codes is from the theoretical limits, we extend the concept of Shannon's sphere packing lower bound (SPLB).
Through our analysis, we are able to identify which are the critical values of the signal-to-interference ratio (SIR) for which the coding scheme is no more able to guarantee an acceptable level of protection.

As further possible candidate schemes, we consider short parallel turbo codes (PTC) and extended BCH (eBCH) codes with soft-decision decoding.
For decoding the eBCH codes we consider the most reliable basis (MRB) algorithm, which has been successfully applied to these codes over the AWGN channel.
We extend its use also to the jamming channel.

The organization of the paper is as follows. In Section \ref{sec:two} we introduce the types of jamming. In Section \ref{sec:three} we give examples of the performance of the BCH$(63, 56)$ code over the jamming channel. In Section \ref{sec:four} we describe the new coding schemes and in Section \ref{sec:five} the performance metrics adopted, including an extension of the SPLB. In Section \ref{sec:six} we provide some numerical examples. In Section \ref{sec:seven} we evaluate the impact of finite length interleavers for the system using PTC. Finally, Section \ref{sec:eight} concludes the paper.

\section{Types of jamming}
\label{sec:two}

The definition of the types of jamming we consider is shortly reminded next. We assume that the system adopts a direct sequence-spread spectrum (DS-SS), and binary phase shift keying (BSPK) modulation with carrier circular frequency $\omega_0$. The DS-SS is characterized by bandwidth $W_{ss}$ and processing gain $K$. Moreover, the length (period) of the spreading sequence is denoted by $L$.

\subsection{Pulsed jamming}
\label{sec:twoA}
A pulsed jamming signal has the following characteristics:
\begin{itemize}
\item white Gaussian noise on the whole bandwidth $W_{ss}$;
\item discontinuity, with pulse active time $D$ and period $T$, which means that the pulse is active for a fraction of time (also
called duty cycle) $0 < \rho = \frac{D}{T} \leq 1$;
\item power $J_P$ during the active time $D$, and zero for the remaining time $T - D$.
\end{itemize}
During the active time the jamming signal has a power spectral density which is constant over the $W_{ss}$ band
with value $\frac{J_{0P}}{2}$, where $J_{0P} = \frac{J_P}{W_{ss}}$. For proper comparison it is also useful to introduce an equivalent (with the same energy) Gaussian continuous
jamming signal. Since the same energy is transmitted over $T$ instead of $D$, it has a power $J = \rho J_P$. This
equivalent jamming signal has a power spectral density constant over the $W_{ss}$ band, with value $\frac{J_0}{2}$, where $J_0 = \frac{J}{W_{ss}} = \rho J_{0P}$. The error rate performance can be expressed in the terms of the ratio $\frac{E_b}{J_0}$ between the energy per bit and the equivalent one-side jamming spectral density.

When an error correcting code is applied, the impact of pulsed jamming can be mitigated through interleaving. This is because most of the forward error correction schemes are designed for an AWGN channel which exhibits no memory. They do not handle bursts of errors. An interleaver distributes a burst of errors among many consecutive codewords. By doing this the number of errors contained in each codeword is limited, the code is able to correct them and the burst is neutralized. For analysis purposes, we can initially refer to an ideal interleaver. This implies that if a burst of errors corresponds to a
fraction $\rho$ of the symbols, its impact after de-interleaving is modeled as a probability  $\rho$, for each symbol, of having a higher noise variance. Even if an ideal interleaver cannot be implemented, it is very useful for analytically investigating the performance over jamming channels. Then, the performance in the presence of a real interleaver can be determined through simulation. 

\subsection{CW jamming}
\label{sec:twoB}
A CW jamming is a narrowband, continuous signal of type
\begin{equation}
j(t) = \sqrt{2J} \cos\left(\omega_j t + \theta_j \right).
\label{CW}
\end{equation}
Hence, it is a pure tone with:
\begin{itemize}
\item circular frequency $\omega_j = 2 \pi f_j$ which, in general, may be different from the signal circular frequency $\omega_0$;
\item initial phase $\theta_j$ which, in general, may be different from the signal initial phase (conventionally set to $0$);
\item power $J$.
\end{itemize}
The worst case occurs when $\Delta \omega = \omega_j - \omega_0 = 0$. Under the hypothesis of having a large $K < L$, the jamming contribution on a generic symbol can be modeled by
a Gaussian random variable with zero mean and variance $\frac{JT_b}{K} \cos^2 \left( \theta_j \right)$, where $T_b$ is the bit time duration. In the case of CW jamming it is preferable to express the error rate performance in terms of the SIR, $\frac{S}{J}$, where $S$ is the signal power. 

\subsection{PN jamming}
\label{sec:twoC}
Let us denote by $c(t)$ the spreading sequence used in the DS-SS system.
A PN jamming is a signal of type
\begin{equation}
j(t) = \sqrt{2J} c_2(t - \tau_2) \cos\left(\omega_0 t \right).
\label{PN}
\end{equation}
Hence, it is a DS-SS signal with:
\begin{itemize}
\item circular frequency equal to $\omega_0$;
\item spreading sequence $c_2(t)$ different from $c(t)$ (although it may have the same length);
\item time delay $\tau_2$ on the spreading sequence;
\item power $J$.
\end{itemize}

A common choice for the spreading sequence $c(t)$ is to use a Gold code. The novel cryptographic PN sequences proposed for TC applications \cite{{Thales2011}} are very long (a suggested length is $L = 2^{22} - 1$). Once again, under the hypothesis of a large $K < L$, the interfering contribution on a generic symbol can be modeled by
a Gaussian random variable with zero mean and variance $\frac{J}{S}\frac{E_b}{K}$. Thus we can obtain the same expression as for CW jamming with $\theta_j = 0$.

For all the types of jamming the channel is also impaired by thermal noise with signal-to-noise ratio per bit $\frac{E_b}{N_0}$. As the two disturbances are independent one each other, when simultaneously present, their variances can be summed. 

\section{Current standard}
\label{sec:three}

\begin{figure}[tb]
\begin{centering}
\includegraphics[width=90mm,keepaspectratio]{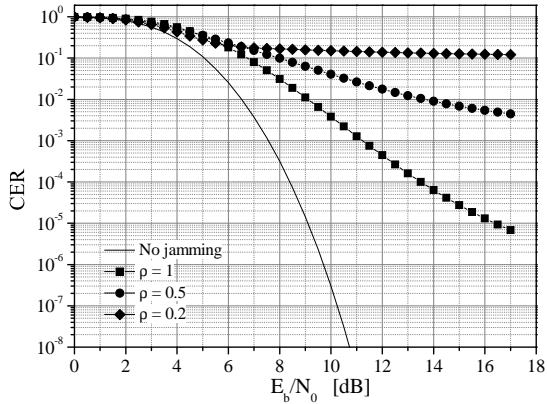}
\caption{CER performance of the BCH$(63, 56)$ code over pulsed jamming channel, for $E_b/J_0 = 10$ dB and no
interleaver. \label{fig:standard_no_int}}
\par\end{centering}
\end{figure}

\begin{figure}[tb]
\begin{centering}
\includegraphics[width=90mm,keepaspectratio]{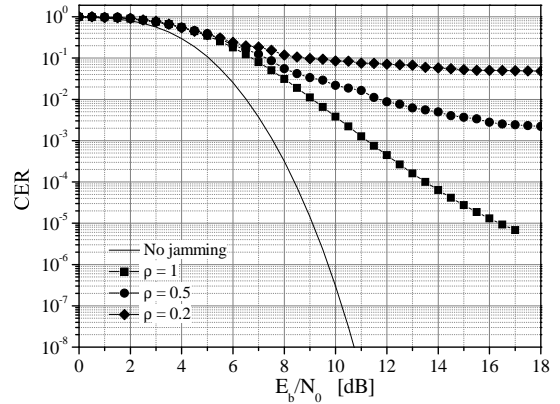}
\caption{CER performance of the BCH$(63, 56)$ code over pulsed jamming channel, for $E_b/J_0 = 10$ dB and ideal
interleaver. \label{fig:standard_with_int}}
\par\end{centering}
\end{figure}

The TC protocols for synchronization and channel coding are specified (with some differences) both in the recommendation \cite{CCSDS2010} issued by the Consultative Committee for Space Data Systems (CCSDS) and in the standard \cite{ECSS2008} issued by the European Cooperation for Space Standardization (ECSS). Let us refer to the CCSDS recommendation \cite{CCSDS2010}: it specifies the functions performed in the ``Synchronization and Channel Coding sublayer'' in TC ground-to-space (or space-to-space) communication links. In short, the sublayer takes transfer frames (TFs) produced by the upper sublayer (``Data Link Protocol sublayer''), elaborates them and outputs Communications Link Transmission Units (CLTUs) that are passed to the lower layer (`Physical layer'') where they are mapped into the transmitted waveform by adopting a proper modulation format. Details on the structure of the TF and CLTU can be found in \cite{CCSDS2010}, \cite{ECSS2008} and are here omitted for the sake of brevity.
The current CCSDS recommendation and ECSS standard use a BCH$(63, 56)$ code for error protection against noise and interference. At the receiver side a hard decision is taken on the received symbols.
The performance against pulsed jamming of the hard-decision decoded BCH$(63, 56)$ code is rather poor. Examples are shown in Fig. \ref{fig:standard_no_int} for the case without the interleaver (as addressed by the current standard) and in Fig. \ref{fig:standard_with_int} for the case with the (ideal) interleaver. Performance is expressed in terms of the CER; the value of $E_b/J_0$ has been set equal to $10$ dB and the CER is plotted as a function of $E_b/N_0$ for some values of $\rho$. As expected, performance degrades for decreasing $\rho$; the use of the (ideal) interleaver introduces for $\rho < 1$ an improvement that however remains unsatisfactory to the point that reaching CER = $10^{-4}$, that is a reference value for TC applications, is practically impossible for $\rho = 0.2$ and $\rho = 0.5$.

\begin{figure}[tb]
\begin{centering}
\includegraphics[width=90mm,keepaspectratio]{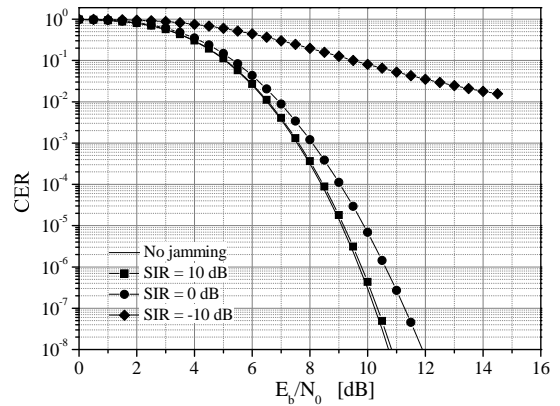}
\caption{CER performance of the BCH$(63, 56)$ code over CW jamming channel ($\theta_j = 0$ and $\Delta \omega = 0$), for $K = 100$. \label{fig:standard_CW}}
\par\end{centering}
\end{figure}

An example of the performance of the hard-decision decoded BCH$(63, 56)$ code against CW jamming is shown in Fig. \ref{fig:standard_CW} for $K = 100$, $\theta_j = 0$ and $\Delta \omega = 0$ (worst case). The SIR value is assumed as a parameter and we see that a SIR in the order of $- 10$ dB makes the system unpractical. A further reduction in the value of $K$ would require higher SIR values; for example, by assuming $K = 10$, the CER target of $10^{-4}$ becomes practically unreachable just for SIR $= 0$ dB.

\begin{figure}[tb]
\begin{centering}
\includegraphics[width=90mm,keepaspectratio]{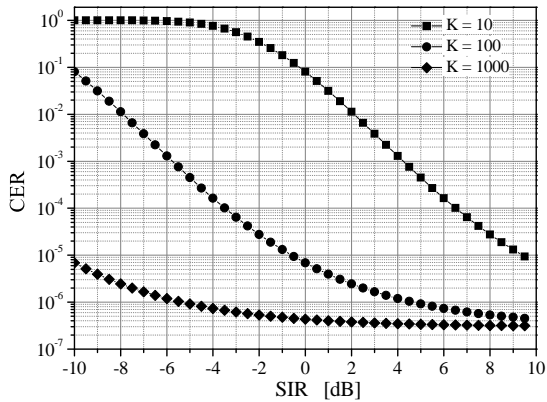}
\caption{CER performance of the BCH$(63, 56)$ code over PN jamming channel, for fixed $E_b/N_0 = 10$ dB. \label{fig:standard_PN}}
\par\end{centering}
\end{figure}

For PN jamming it is possible to verify that under the Gaussian approximation its impact is equivalent to that of CW jamming with $\theta_j = 0$ and $\Delta \omega = 0$. In Fig. \ref{fig:standard_PN} we have plotted the CER curve as a function of the SIR for different values of $K$ and $E_b/N_0 = 10$ dB. So this curve also applies to the worst case CW jamming with the same SIR.

\section{New coding schemes}
\label{sec:four}
The codes proposed for TC channel coding updating have length greater ($n = 128, 256$ and $512$) and rate smaller ($1/2$) than the standard code. Also the code type has been changed, with the aim to introduce state-of-the-art codes. The mostly addressed candidates, in this sense, are LDPC codes, both binary and non-binary. Recently, however, we have also shown that potential competitors can be PTC and even BCH codes, if they are soft-decision decoded with maximum likelihood (ML)-like algorithms characterized by limited complexity.
The main features of these schemes are reminded next.

\subsection{LDPC codes}
\label{sec:four_a}

A class of binary LDPC codes that is suitable for TC applications has been proposed by the National Aeronautics and Space Administration (NASA) and is described in \cite{CCSDS2012}. It is based on the adoption of three systematic short binary LDPC codes designed using protographs with circulant matrices. Soft-decision decoding can be realized by using the classic sum-product algorithm with log-likelihood ratios (LLR-SPA).

Non-binary LDPC codes with the same lengths have been analyzed by NASA \cite{NASA/JPL2012} and independently by the Deutsches Zentrum f\"{u}r Luft- und Raumfahrt (DLR) in a joint work with the University of Bologna (UniBO) \cite{Costantini2012}. In this paper we refer to the DLR-UniBO implementation. Decoding is realized by using iterative algorithms based on fast Hadamard transforms.

\subsection{Parallel turbo codes}
\label{sec:four_b}
Parallel turbo codes are one of the coding options of the CCSDS recommendation for telemetry (TM) links \cite{CCSDS2011}. The CCSDS turbo encoder is based on the parallel concatenation of two equal 16-state systematic convolutional encoders with polynomial description  $(1, (1+D^2 + D^4 + D^5)/(1 + D^3 + D^4))$. The interleavers are based on an algorithmic rule proposed by Berrou and described in \cite{CCSDS2011}. The CCSDS turbo encoder has four possible information frame lengths: $1784, 3568, 7136$ and $8920$ bits. The nominal code rate can be $1/2, 1/3, 1/4$ and $1/6$. However, higher rates are obtainable by puncturing \cite{Calzolari2005}.

Maintaining unchanged the encoder structure, we have considered frame lengths shorter than those in the TM recommendation and fixed the nominal code rate to $1/2$, in such a way as to comply with the NASA's choices discussed above.
Because of the shorter length, we cannot use the interleavers in \cite{CCSDS2011} and we must design new smaller interleaving structures.

Among a wide number of different options, we have focused attention on: completely random, spread \cite{Divsalar2005}, Quadratic Permutation Polynomial (QPP) \cite{Sun2005} and Dithered Relative Prime (DRP) \cite{Croizer2001} interleavers.
Moreover, since the constituent CCSDS convolutional codes have $16$ states, four extra-tail bits are needed for termination; then, the turbo codeword length is $n = 2(k + 4)$. As an example, for the case of $k = 64$, this implies to have $n = 136$ and an actual code rate $0.471$. 
In order to achieve exactly the code rate $1/2$, as it is necessary for fair comparison, we have implemented a suitable puncturing strategy.

The interleaver and the puncturing pattern have been jointly optimized, in such a way as to maximize the minimum distance $d_{min}$ and minimize the codewords multiplicity $A_{min}$ (i.e., the number of codewords with Hamming weight $d_{min}$); these parameters, in fact dominate the code performance at low error rates.

The results of the design optimization are shown in Table \ref{tab:syst}; besides $d_{min}$ and $A_{min}$, also the information multiplicity $w_{min}$ (sum of the input weights over all the codewords with weight $d_{min}$) is provided since, together with $d_{min}$, it determines the asymptotic bit error rate performance.
\begin{table}[tb]
\caption{Selected interleavers for the parallel turbo codes.}
\scriptsize
\begin{center}
\begin{tabular}{c|c|c}
Input length $k$ & Interleaver & Minimum distance parameters\\
 & & $d_{\min}/A_{\min}/w_{\min}$ \\
\hline
\hline
64 & DRP & 10/5/23 \\
\hline
128 & DRP & 13/13/40 \\
\hline
256 & QPP & 15/1/1 \\
\hline
\end{tabular}
\end{center}
\label{tab:syst}
\end{table}
The decoding of turbo codes is performed by iteratively applying the well-known BCJR algorithm \cite{Bahl1974} to the constituent
encoders.

\subsection{Extended BCH codes}
\label{sec:four_c}

Although soft-decision decoding of BCH codes is generally complex, in the case of short BCH codes with high rate, an exact ML soft-decision decoding is possible, through its  trellis representation (for example, based on the Viterbi or the BCJR algorithms). However, having now decided to use codes with rate $1/2$ and the shortest length $128$, these techniques are too involved and therefore cannot be applied. Alternative solutions can be found, at least for the case of $k = 64$.
In fact, the eBCH$(128, 64)$ code can be efficiently decoded by using sub-optimal soft-decision decoding algorithms.
Several options are available for this purpose, also exploiting LDPC-like code representations \cite{Baldi2007}.
For this code, we have focused on the MRB algorithm \cite{Wu2007}, which has very good performance and acceptable complexity.
It consists of the following steps:
\begin{enumerate}
\item Identify $ k= 64$ most reliable received bits and obtain from them a vector $v^*$.
\item Construct a systematic generator matrix $G^*$ corresponding to these bits.
\item Encode $v^*$ by $G^*$ to obtain a candidate codeword $c^* = v^*G^*$.
\item Choose the order $i$ of the algorithm.
\item Consider all (or a proper subset of) test error patterns of length $k$ and weight $w \leq i$.
\item For each of them: sum to $v^*$, encode by $G^*$, verify if the likelihood is higher than that of the previous candidate codeword and, if this is true, update the candidate.
\end{enumerate}
Further details can be found in \cite{Wu2007} and the references therein. The MRB algorithm can be applied also to the other schemes (to LDPC codes, in particular) where, depending on the order $i$, it can provide performance comparable to, or even better than, that offered by the iterative algorithms.

\section{Performance evaluation}
\label{sec:five}

The performances of the new coding schemes presented in Section \ref{sec:four} have been recently discussed and compared over the AWGN channel \cite{Baldi_VTC2013}. We have verified they can provide an advantage of more than $5$ dB over the current BCH$(63, 56)$ code. In Section \ref{sec:six} we will show that similar improvements can be achieved against jamming.

For soft-decision decoding, the knowledge of the jamming state can play a relevant role. For the case of pulsed jamming, for example, to have JSI means to know the noise variance for each symbol. Thus, in our simulations, we have considered the case of perfect JSI, which means the receiver is able to identify the fraction, $\rho$, of symbols affected by jamming and the remaining fraction, $1 - \rho$, that is only affected by thermal noise, and properly estimate their noise variances, which are $\sigma_\rho^2 = \frac{N_0}{2} + \frac{J_0}{2\rho}$ and $\sigma_{1-\rho}^2 = \frac{N_0}{2}$, respectively. On the opposite side, we have also considered the case when JSI is not available, and the variance used for LLR calculation of the decoder input is always equal to the average value $\frac{N_0}{2} + \frac{J_0}{2}$.

The goodness of the proposed solutions can be measured through the distance of the CER curves from the SPLB. Among the various approaches available to compute the SPLB, the most suitable one is the so-called SP59 \cite{Shannon1959}. A modified version of this bound is also available (called SP67 \cite{Shannon1967}), that is able to take into account the constraint put by the signal constellation (BPSK in the present analysis). More recent improvements \cite{Valembois2004}, \cite{Wiechman2008} are significant only for high code rates or long codeword lengths, and these conditions are not satisfied by the codes here of interest. Thus, in the present study, we consider the SP59 as the most significant SPLB.

The SPLB reported in the literature refers to the AWGN channel and needs generalization. This is easy to achieve for pulsed jamming on the condition that the pulse duration is a multiple of the codeword length and no interleaving is applied.
For this purpose, let us denote by SPLB$(R_c,n,E_b/N_0)$ the SP59 bound for the AWGN channel. Under the hypotheses above an extended sphere packing lower bound, ESPLB, for the case of pulsed jamming can be defined as:
\begin{eqnarray}
\label{eq:ESPLB}
CER & \geq & ESPLB \left ( R_c, n, \frac{E_b}{N_0}, \frac{E_b}{J_0}, \rho \right ) \nonumber \\ 
& = & \rho SPLB \left ( R_c, n, \frac{1}{\frac{1}{\frac{E_b}{N_0}} + \frac{1}{\frac{E_b}{J_0}\rho}} \right ) \\
& + & (1 - \rho) SPLB \left ( R_c, n, \frac{E_b}{N_0} \right ). \nonumber
\end{eqnarray}
By setting $\rho = 1$ in \eqref{eq:ESPLB}, we obtain an expression which is valid also for CW and PN jamming channels, when the Gaussian approximation is applied.
The ESPLB given by \eqref{eq:ESPLB} will be considered in Section \ref{sec:six} as a useful benchmark for the case with JSI and without interleaving. 

\section{Numerical examples}
\label{sec:six}
Due to limited space, we focus on pulsed jamming and on codes with $n = 128$ and $k = 64$. The analysis can be extended to the longer codes for which, however, the complexity issue for the decoding algorithms adopted can become more critical.

\begin{figure}[tb]
\begin{centering}
\includegraphics[width=90mm,keepaspectratio]{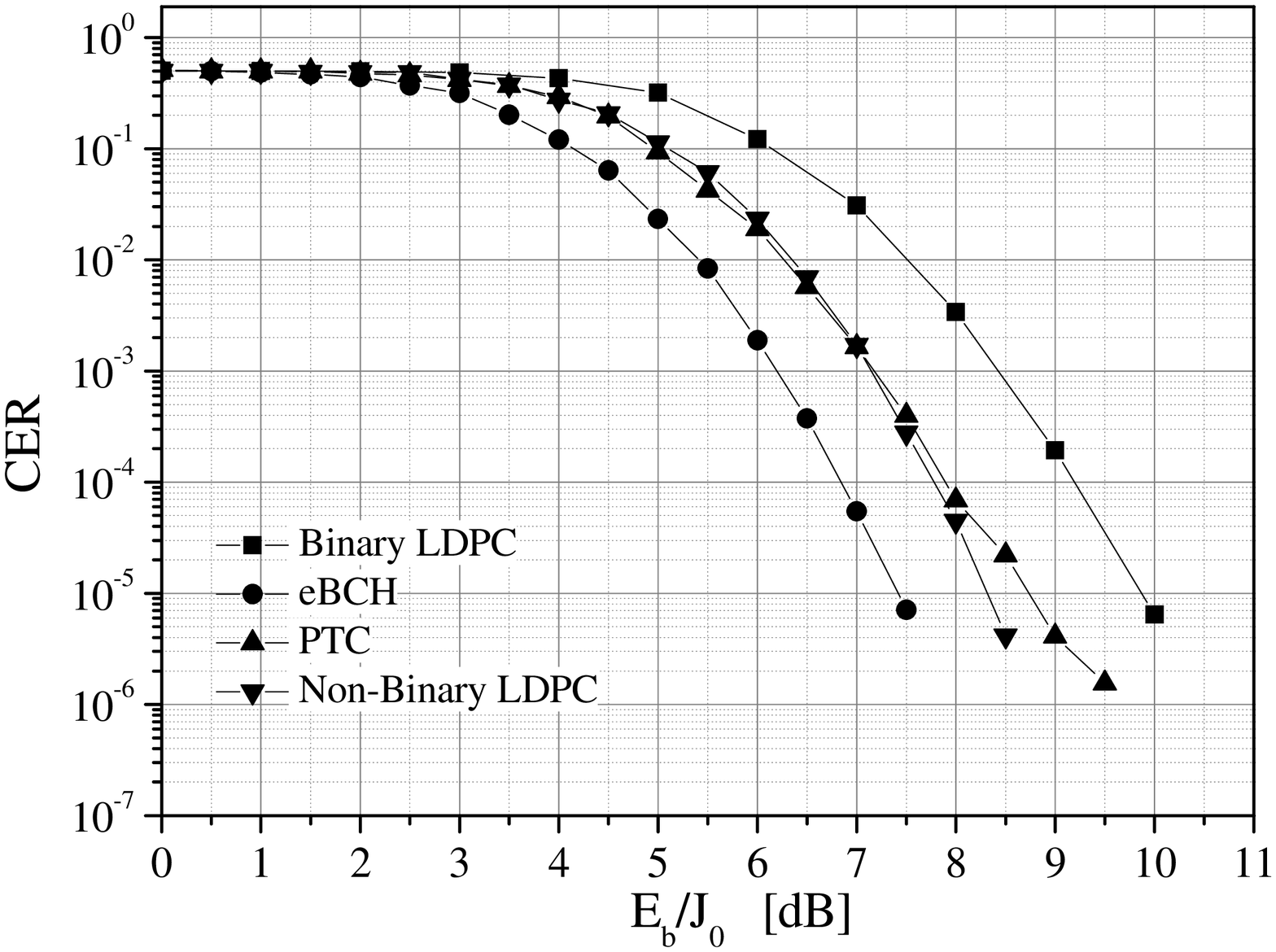}
\caption{Performance of the new coding schemes ($n = 128, k = 64)$  without interleaving and without JSI. \label{fig:Results_noINT_noJSI}}
\par\end{centering}
\end{figure}

\begin{figure}[tb]
\begin{centering}
\includegraphics[width=90mm,keepaspectratio]{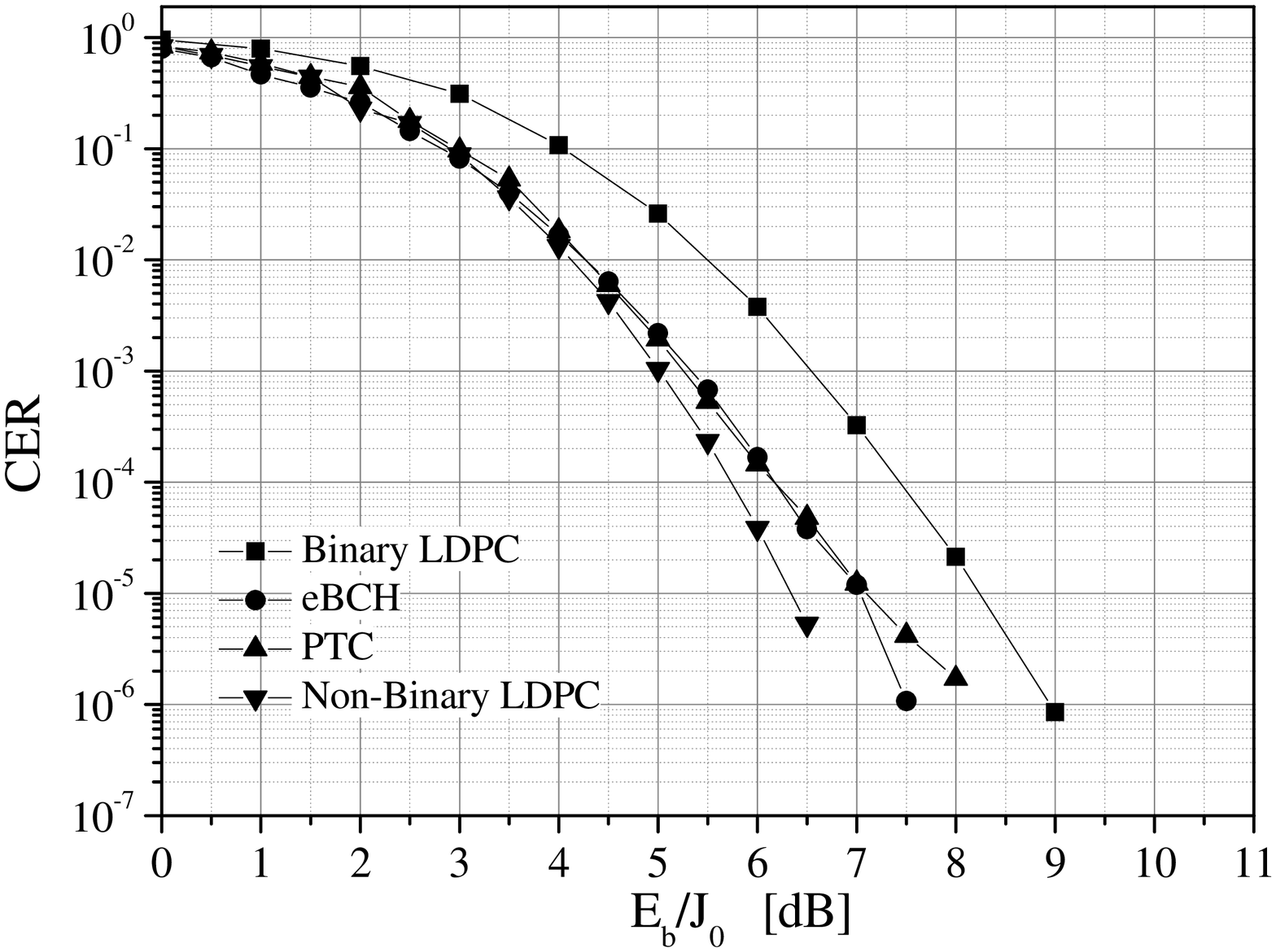}
\caption{Performance of the new coding schemes $(n = 128, k = 64)$ with interleaving and without JSI. \label{fig:Results_withINT_noJSI}}
\par\end{centering}
\end{figure}

\begin{figure}[tb]
\begin{centering}
\includegraphics[width=90mm,keepaspectratio]{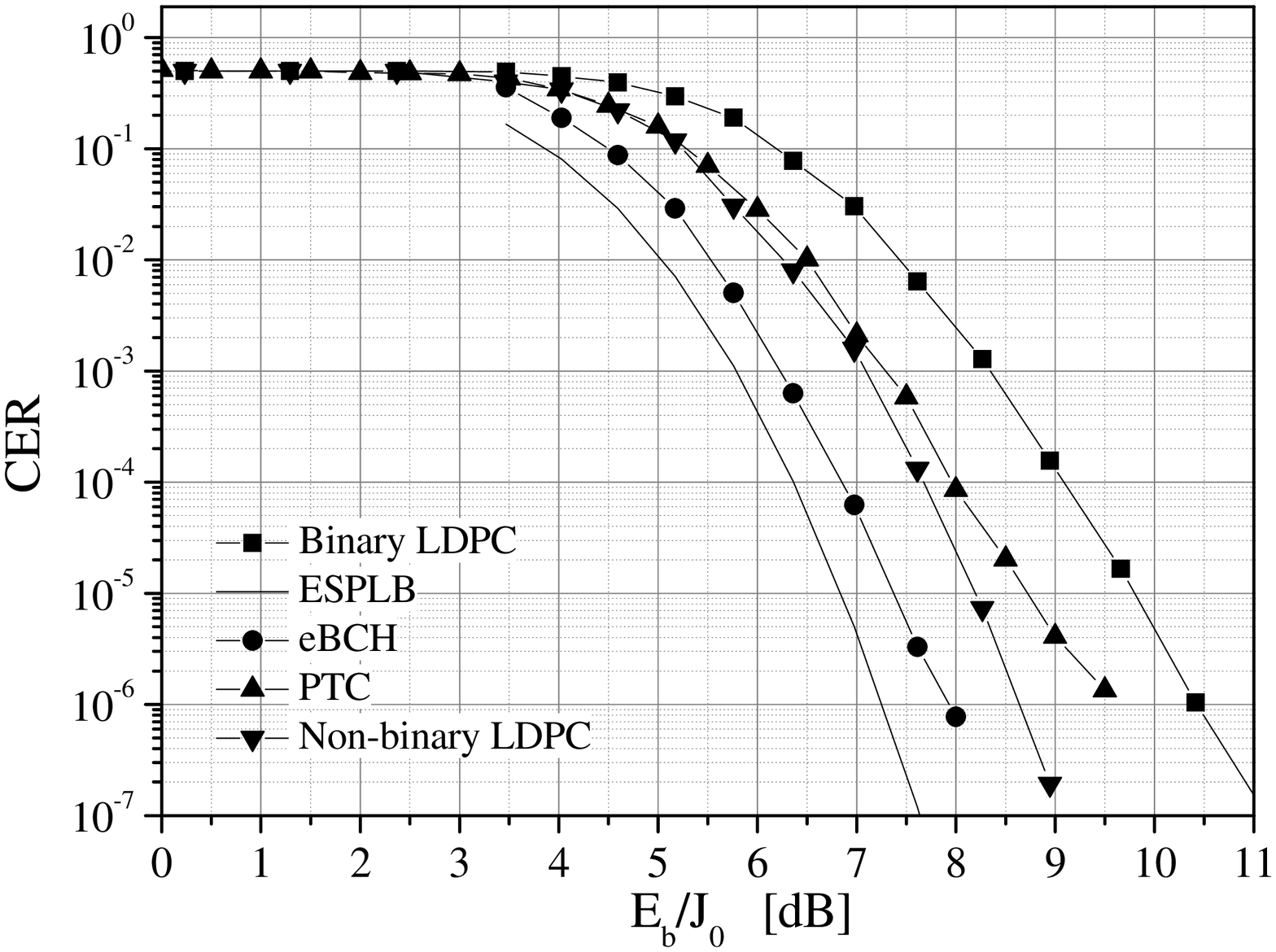}
\caption{Performance of the new coding schemes $(n = 128, k = 64)$ without interleaving and with JSI. \label{fig:Results_noINT_withJSI}}
\par\end{centering}
\end{figure}

\begin{figure}[tb]
\begin{centering}
\includegraphics[width=90mm,keepaspectratio]{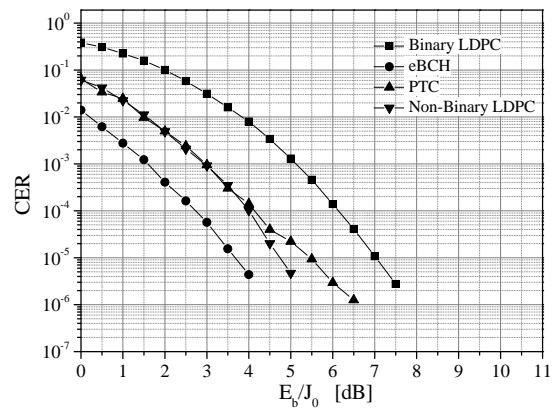}
\caption{Performance of the new coding schemes $(n = 128, k = 64)$ with interleaving and with JSI. \label{fig:Results_withINT_withJSI}}
\par\end{centering}
\end{figure}

The performances of the new coding schemes presented in Section \ref{sec:four} are compared in Figs. \ref{fig:Results_noINT_noJSI}-\ref{fig:Results_withINT_withJSI} assuming $\rho = 0.5$, $E_b/N_0 = 10$ dB and variable $E_b/J_0$, with and without an (ideal) interleaver, with and without JSI. In the latter case, to limit the impact of the incorrect noise estimation a clipping threshold, equal to twice the amplitude, has been applied to the signal at the channel output. For the PTC we have used the optimal DRP interleaver reported in Table \ref{tab:syst}. The order used for the MRB algorithm applied to the eBCH code is $i = 4$.

For three of the considered scenarios the relative behavior of the proposed schemes is very similar: the best performance is achieved by the eBCH code, while PTC and non-binary LDPC codes are very close one each other and suffer a penalty with respect to the eBCH code that depends on the simulation conditions. An interesting exception occurs for the case with interleaving and without JSI where, because of the ordering mechanism that is at the basis of MRB, the eBCH code loses its leadership.
The eBCH code is also very close to the ESPLB, where applicable (see Fig. \ref{fig:Results_noINT_withJSI}). On the contrary, the performance of the binary LDPC code is rather poor with a loss than can be larger than $3$ dB with respect to the best solution. These gaps are even more pronounced than those found over the AWGN channel \cite{Baldi_VTC2013}.

\section{Impact of finite length interleaver}
\label{sec:seven}
The results shown in the previous section for the systems using interleaving referred to the adoption of an ideal interleaver. In this section we  discuss the effect of using real interleavers characterized by finite length. We consider the particular case of using PTC, but a similar analysis could be developed for the other schemes.

\begin{figure}[tb]
\begin{centering}
\includegraphics[width=90mm,keepaspectratio]{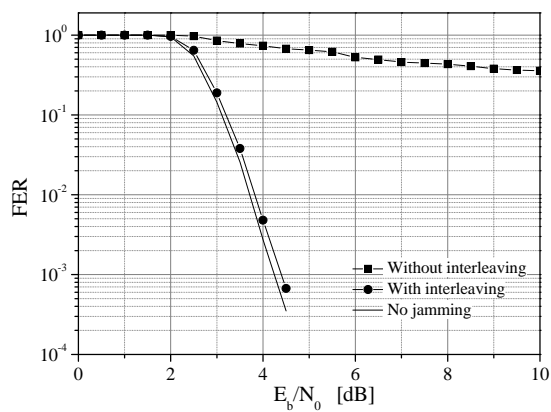}
\caption{Performance of the PTC$(128, 64)$ without interleaving and using a $64 \times 64$ row-by-column interleaver, for a burst length of $100$ bits; ${E_b}/{J_{0P}} = 0$ dB and $M = 2048$. \label{fig:real_int}}
\par\end{centering}
\end{figure}

With reference to the short description given in Section \ref{sec:three} (but further details can be found in \cite{CCSDS2010}), we suppose that interleaving is applied at CLTU level and also taking into account the possible presence of partitioning. This occurs when the TF has length $M$ that is not a multiple of $k$: the TF is partitioned into $N = \lceil \frac{M}{k} \rceil$ input blocks and, if needed, zero filling is used to complete the last block. Each block is then encoded producing $C = Nn$ CLTU coded bits. For the sake of simplicity, in this first evaluation we neglect the presence of the preamble ($16$ bits) and the postamble (64 bits) that are added for CLTU synchronization \cite{CCSDS2010}. So, we assume that an interleaver is applied to the $C$ CLTU coded bits to increase the protection against bursts; it involves all the $N$ codewords of the CLTU. As a simple example, we consider a square $R \times R$ row-by-column interleaver where $R = \lceil \sqrt{C} \rceil$; the $C$ CLTU coded bits are written by row and read by column (or vice versa). In this case, wishing to compare the impact of bursts of growing length it is preferable to refer directly to the power $J_P$ and the corresponding ratio $\frac{E_b}{J_{0P}} = \frac{S}{J_{P}}K$. An example of the impact of a real interleaver on the transfer frame error rate (FER) in the considered scenario is shown in Fig. \ref{fig:real_int}, for $\frac{E_b}{J_{0P}} = 0$ dB, $M = 2048$ and a burst length of $100$ bits. We observe that the CLTU interleaver is very effective against bursts produced by pulsed jamming. For long TFs (like the one here considered) the burst is practically neutralized by the interleaver.

\section{Conclusion and open issues}
\label{sec:eight}
For the first time at our knowledge, the performance over jamming channels of new coding schemes that potentially replacing the BCH$(63, 56)$ code in the space TC channel coding standard has been investigated. The adoption of soft-decision decoding, combined with interleaving and JSI, allows to achieve significant improvements. Whilst it is confirmed that non-binary LDPC codes, that are considered the most eligible candidates, are generally very good, we have shown that comparable performances can also be achieved by using PTC or eBCH codes. We have studied the case of $n = 128$ and $k = 64$, where eBCH codes often offer the best results at an acceptable complexity.

This fact, however, cannot be used to draw general conclusions. First of all, extending the sub-optimal decoding algorithms to longer codes, while maintaining acceptable complexity, may be quite difficult. Additionally, the sub-optimal algorithms generally define ``complete'' decoders. As it is well known, this may be a penalty for the undetected codeword (or frame) error rate that, in TC applications, is at least as important as the codeword (or frame) error rate. Further work is progress to assess either the complexity and the completeness issues.

\newcommand{\BIBdecl}{\setlength{\itemsep}{0.01\baselineskip}}

\end{document}